\documentclass[a4paper,11pt]{article}
\usepackage{aaskaiid}
\usepackage{orcidlink}

\newcommand{\msun}{$\rm M_{\odot}$ }

\title{Citizen Science Research with the Square Kilometre Array Observatory (SKAO)}
\ShortTitle{Citizen Science Research with the SKAO}

\author[1,2,3]{Ananda Hota\orcidlink{0000-0002-4959-7376}}
\ShortName{Ananda Hota et al.} 
\author[4,3]{Pratik Dabhade\orcidlink{0000-0001-9212-3574}}
\author[5]{M. J. Hardcastle\orcidlink{0000-0003-4223-1117}}
\author[6]{Stephen Serjeant\orcidlink{0000-0002-0517-7943}}
\author[7,3]{C. Konar\orcidlink{0000-0002-2530-3812}}
\author[6]{Hugh Dickinson\orcidlink{0000-0003-0475-008X}}
\author[8,3]{Sabyasachi Pal\orcidlink{0000-0003-2325-8509}}
\author[3]{Sravani Vaddi\orcidlink{0000-0003-3295-6595}}
\author[9,10,3]{Sagar Sethi\orcidlink{0000-0001-8561-4228}}
\author[11,12,3]{Shubhrangshu Ghosh}
\author[3]{Arundhati Purohit}

\affiliation[1]{UM-DAE Centre for Excellence in Basic Sciences, University of Mumbai, Santacruz-East, Mumbai400098, India }
\emailAdd{hotaananda@gmail.com}
\affiliation[2]{Centre for Excellence in Theoretical and Computational Science, University of Mumbai, Santacruz-East, Mumbai400098, India}
\affiliation[3]{RAD@home Astronomy Collaboratory, Kharghar, Navi Mumbai PIN 410210, India}
\affiliation[4]{Astrophysics Division, National Centre for Nuclear Research, Pasteura 7, 02-093 Warsaw, Poland}
\affiliation[5]{Centre for Astrophysics Research, Department of Physics, Astronomy and Mathematics, University of Hertfordshire, College Lane, Hatfield AL10 9AB, UK}
\affiliation[6]{School of Physical Sciences, The Open University, Walton Hall, Milton Keynes, MK7 6AA, UK} 
\affiliation[7]{Department of Physics, Amity Institute of Applied Sciences, Amity University Uttar Pradesh, Sector-125, Noida 201313, U.P., India}
\affiliation[8]{Department of Pure and Applied Sciences, Midnapore City College, Kuturia, Bhadutala, Paschim Medinipur, West Bengal, 721129, India}
\affiliation[9]{Space Radio-Diagnostics Research Centre, University of Warmia and Mazury, R. Prawochenskiego 9, 10-719 Olsztyn, Poland}
\affiliation[10]{Astronomical Observatory, Jagiellonian University, ul. Orla 171, 30-244 Kraków, Poland}

\affiliation[11]{Center for Astrophysics, Gravitation and Cosmology, Shri Ramasamy Memorial (SRM) University Sikkim, 5th Mile Tadong, Gangtok, 737102, India}
\affiliation[12]{Department of Physics, Shri Ramasamy Memorial (SRM) University Sikkim, 5th Mile Tadong, Gangtok, 737102, India}

\abstract{
Over the past two decades, internet-enabled citizen science research (CSR) has contributed to significant discoveries while involving millions of people in the research process. Our review highlights CSR in extragalactic radio astronomy and emphasises that such approaches will become increasingly relevant across radio astronomy in the era of the Square Kilometre Array (SKA). As astronomical data volumes grow, CSR is converging with Artificial Intelligence and Machine Learning (AI/ML), creating hybrid human–machine frameworks suited to face big-data challenges. Two CSR platforms, such as Radio Galaxy Zoo and RAD@home, demonstrate success: the former excels in large-scale, web-based catalogue creation, while the latter combines structured training with collaborative discovery. Following this, we propose CSR with the SKA, SKA@home,  with two modes, one purely web-based and the other in collaboratory mode with national training programmes. We argue that CSR can complement, and at times surpass, automated AI/ML pipelines, particularly in identifying rare, intricate, or unexpected features. Illustrative CSR discoveries include an episodic wide-angle-tailed radio galaxy, a jet–galaxy interaction, a collimated synchrotron thread, a twin-ring odd radio circle, and a large-scale shock ahead of a cluster-infalling galaxy. Consistent with the IAU’s recognition of CSR as a driver of Astronomy for Development and the United Nations’ affirmation of participation in science as a universal human right, both the SKA construction proposal and outreach strategies show commitments to enabling CSR with SKA. Proposed SKA@home would not only enhance early discovery potential of SKA data but also initiate deeper meaningful connection with the larger society. 
}

\begin{document}
\maketitle



\section{Introduction}\label{sec:Intro}
The pursuit of knowledge and the advancement of research have historically been primarily driven by individual curiosity and intellectual passion. In earlier eras, scholars who engaged in the generation and dissemination of new knowledge were often supported through royal patronage. Over time, this scholarly class evolved into professional educators and scientists employed within academic and research institutions funded by government or public bodies. Prior to the institutionalisation and professionalisation of scientific inquiry and teaching, the pursuit of science was largely undertaken by individuals who, in essence, were amateur researchers motivated by a profound commitment to understanding the natural world.

In astronomy, the term amateur astronomer remains a respected and popular informal designation, unlike in many other branches of science. These individuals observe the Sun by day and other celestial objects by night in their quest for knowledge and to satisfy intellectual curiosity. Some amateurs develop considerable expertise and acquire their own equipment (telescopes) that is capable of contributing valuable data to international research publications and databases \cite[]{AmateurContribution-Romanov2022}with. This enduring tradition of voluntary, high-quality contributions to knowledge creation, recognised by professional astronomers, is being redefined in modern times as `citizen science', which is synonymous with terms like community science, crowd science, crowd-sourced science, and civic science  \citep{CSR-Irwin1995_CitizenScience, CSR-Haklay2021,10principles-CSR-Robinson2018, CSR-benefit-challenges-Franzoni2014, CSR-book-Nielsen}. In this chapter, we will refer to it as Citizen Science Research (CSR).

With time, telescopes became too big for individuals to build and own. Research or educational institutions built and operated large observatories that are mostly located away from the city. Modern observatories now produce digitised data that can be accessed by anyone via the internet. These observatories conduct all-sky imaging surveys and generate catalogues that continue to be valuable to astronomers worldwide, even after the initial refereed publications by the teams involved in data acquisition. Recognising that a large portion of data collected by publicly funded telescopes and astronomers often remained unpublished, observatories introduced a policy to make such data publicly available after a proprietary period. This open release of previously unpublished data has become a vital resource for underfunded astronomers working outside national observatories. The practice of public data release has evolved into a new and inclusive research culture in astronomy. In some cases, independent researchers have published more papers from the archival data than the Principal Investigators using that telescope \citep{Archive-White2009, NASEM2021Pathways}. This represents a remarkable, open, and collaborative model of scientific progress in the field of astronomy. 

Although astronomical survey data are fully accessible to the general public, only skilled astronomers were initially able to utilise them (mine the data) for serious research publications \citep[]{DataMiningBorne2009}. However, certain types of image analysis can be performed by anyone with basic computer literacy and high-school–level understanding. Simple tasks such as classifying galaxies into spiral, elliptical, or irregular types can easily be carried out by the public following clear online instructions. This approach, engaging non-specialists in pattern recognition and image classification, led to the creation of projects like Galaxy Zoo. In its first galaxy classification study, multi-colour optical images from the Sloan Digital Sky Survey (SDSS) were classified by over 100,000 volunteers worldwide \citep{GalaxyZoo-Lintott2008-paper1}.
 
Since then, nearly three million people have registered on the Zooniverse platform\footnote{\url{https://www.zooniverse.org/}}, which evolved from the Galaxy Zoo astronomy project and now hosts dozens of initiatives across both science and the humanities. The web interface for contributing to Zooniverse - through human pattern recognition of images — is so intuitive that it can be used even on a smartphone with internet access, without the need for a computer. This simplicity embodies the empowering nature of CSR, where data collected by sophisticated world-class instruments are analysed by individuals anywhere in the world, including those in small, resource-limited villages. Such accessibility has been recognised as a great social equaliser. The inclusive and non-discriminatory character of modern CSR, uniting people globally for a common scientific purpose, is unprecedented. Consequently, organisations such as the International Astronomical Union (IAU) regard CSR as a vital component in achieving Astronomy for Development \footnote{\url {https://astro4dev.org/}} and the United Nations Sustainable Development Goals (SDGs)\footnote{\url{https://sdgs.un.org/goals}}. Some scholars have even argued that the right to participate in CSR should be recognised as a human right, ensuring that people at the base of the socio-economic pyramid share in the benefits of scientific progress. Considering all these factors, many new astronomical observatories are incorporating science-ready data archives, astronomical user support centres, education and public outreach (EPO) infrastructures to increase both scientific and societal benefits \citep[]{LSST-EPO-Metzger2023Mirro...4...26M, NASEM2021Pathways, CSR-research-infrastructure-Serjeant2024, CSR-benefit-challenges-Franzoni2014}. Therefore, integrating CSR into future large-scale research facilities — such as the Square Kilometre Array Observatory (SKAO) — funded by taxpayers from the participating countries, should be strategically planned well before commissioning. It can be planned as an EPO infrastructure or through a separate technical/social science working group. Doing so will help foster inclusivity and contribute to building a more equitable and harmonious world.

\section{Citizen Science vs Research-Education-Outreach}
\begin{figure}[h]
    \centering
	\includegraphics[width=0.3\columnwidth]{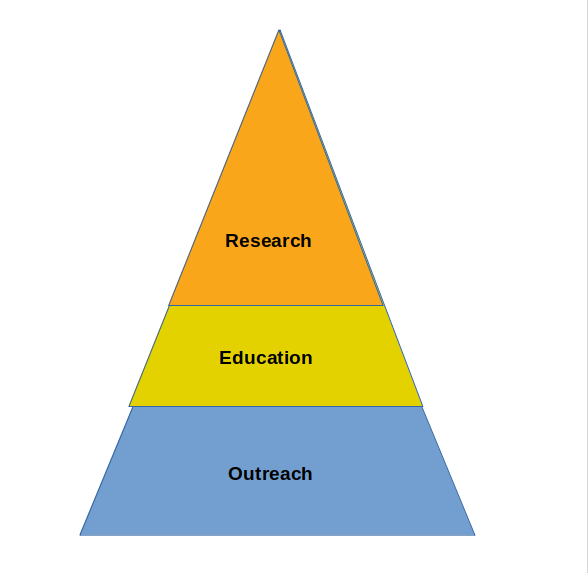}
    \includegraphics[width=0.3\columnwidth]{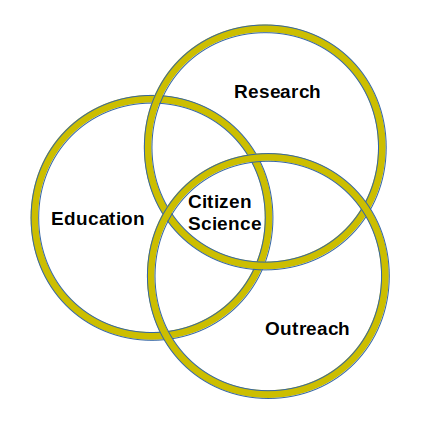}
    \includegraphics[width=0.3\columnwidth]{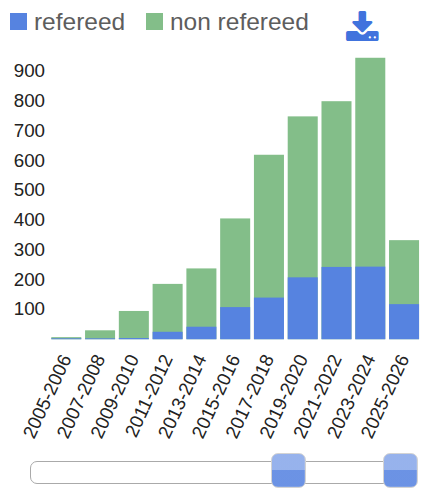}
        \caption{Schematic for the pyramidal structure of the traditional Outreach, Education and Research system based on the number of people involved (left panel). Schematic Venn diagram of modern citizen science, which has significant overlap with all Outreach, Education and Research and large public involvement (middle panel). The image shows how the number of papers mentioning "citizen science" in their abstracts has been growing over the years, plotted since 2005 and as available in the Astrophysics Data System (right panel).}
    \label{fig:fig1}
\end{figure}
The traditional research, education, and outreach system is often visualised as a pyramid, especially due to the number of people involved in it (Fig.~\ref{fig:fig1} left panel). Progress upward through this structure requires increasing levels of effort, expertise, and specialisation, resulting in fewer and fewer people involved. At the broadest base are public lectures and outreach programs, which engage the largest number of participants. These are open to everyone without eligibility criteria, allowing unrestricted public participation. Education forms the next level, involving fewer people due to merit-based selection, admission requirements, a limited number of seats available, and quantitative evaluation before advancement to higher levels. The knowledge gained through education, from any source, can be freely shared, making one a teacher, either as a professional educator earning fees from students or as a salaried instructor in public institutions. At the narrowest top of the pyramid lies research, involving the smallest number of participants and the highest level of expertise. Unlike education, research lacks standardised quantitative measures for comparison between individuals. A researcher’s primary role is to create new knowledge through sustained effort, which must then be carefully documented and peer-reviewed before global publication. Researchers, often supported by public funds, receive salaries and grants to advance this process of knowledge creation. They also develop new instruments and methodologies to enhance investigation and further the frontiers of understanding.

The enormous volume of data generated by modern research facilities has become increasingly difficult to manage by the relatively small number of scientists at the top of the traditional research pyramid \citep{BigData-Bell2009}. With the advent of the internet, it was soon realised that certain simple data analysis tasks — such as well-defined pattern recognition in images- could be distributed to a large number of participants, even at the student level. This approach has enabled many more people to engage directly in the research process, though they may not contribute to every stage required for producing a formal scientific manuscript. Through participation in such web-based interfaces, individuals learn passively yet meaningfully, depending on their curiosity, effort, and the opportunities provided by the project. In this way, modern CSR blends the boundaries between research, education, and outreach (see Fig.~\ref{fig:fig1} middle panel) and involves millions of people, even before they complete University Education, in the research process  \citep{Learning-through-CSR-NAP25183}. News of a scientific discovery can inspire today’s students to become tomorrow’s researchers, and through CSR, those same students may even participate as co-discoverers in scientific breakthroughs. Though their contributions may be modest, this experience helps them grow into more capable and well-rounded scientists through continued formal education and engagement \citep{science-of-CSR-Book-Vohland2021}.

There are inherent limitations in the traditional Research–Education–Outreach system that internet-enabled CSR can largely help to overcome. Although quality research and educational institutions select their members based on merit, they are typically located in urban areas, limiting access for the public in remote regions. Every attempt should be made to address various socio-economic and geo-political inequalities in our access to quality education\footnote{ \url{https://unesdoc.unesco.org/ark:/48223/pf0000373718 }}. To mitigate this, many outreach programs are now organised online, with recorded lectures made freely available. The benefits of Massive Open Online Courses (MOOCs) are undeniable; however, the rigour of such online education and the credibility of their merit certificates are often questioned \citep{Hollands2014MOOC,MOOCS-Laurillard_2016}. Internships and project-based learning have become essential for helping university students transition smoothly from education to research \citep{Internship-Lopatto}. Yet, educational institutions often struggle to provide meaningful projects to large numbers of students due to limited faculty capacity. Similarly, research institutes typically accept only a small number of highly meritorious students for internships involving genuine research participation. CSR can help address this gap if suitably adapted to award participants with merit certificates and recommendation letters \citep{CSR-UG-internship-Mitchell2017, Internship-Lopatto}. Like outreach, if every observatory were to allocate even a small fraction of its telescope time to CSR projects using unpublished data, several limitations of the existing Research–Education–Outreach model could be mitigated, enhancing human resource development across nations \citep{CSR-Bonney2014Science}. 
For instance, the Las Cumbres Observatory program focuses primarily on education and training, though its telescopes are designed for instructional rather than research-grade use \citep{LasCumbres-Brown2013}. The Spitzer Space Telescope Research Program for Teachers and Students exemplifies a world-class research facility that provides data for educational purposes in an organised manner \citep{Spitzer-Spuck2010}. Similarly, the Vera C. Rubin Observatory (formerly the Large Synoptic Survey Telescope) is developing education and public outreach (EPO) programs that will provide support to researchers who create CSR projects using Rubin Observatory data\footnote{\url{https://www.lsst.org/about/epo}}. This EPO activity is fully funded even during the construction phase before any science data release \citep{LSST-EPO-Metzger2023Mirro...4...26M}. A significant growth in the use of the citizen science approach to scientific research publications can be seen over the last two decades. A plot of the number of research papers published/presented over the years, with "citizen science" in its abstract, as available in the Astrophysics Data System\footnote{\url{https://ui.adsabs.harvard.edu}} shows that over 100 research publications every year since the last ten years(Fig.~\ref{fig:fig1} right panel). There is a growing need for all world-class facilities to contribute a portion of their unpublished data to CSR -potentially through qualified CSR-projects which would guarantee its scientific use - to advance education, inclusivity and global participation in research\footnote{\url{https://astro4dev.org/}}.

\section{Various Models of Citizen Science Research}
In other academic disciplines, such as the social sciences, the level of collaboration between volunteers and professional researchers can be so deep as to involve the volunteers at every stage of the ``research cycle" (inasmuch as it's ever cyclic in practice), including involving volunteers in the formulation of the fundamental research questions. This approach is sometimes referred to as "participatory research". However, it is largely impossible in practice to apply this in the physical sciences, because the necessary starting point for having sufficient background to formulate the fundamental research questions is, almost always, a graduate or postgraduate degree. Modern CSR can be broadly classified into four models, distinguished primarily by the nature of the volunteer’s contribution. We elaborate on each of these below with examples.

\textbf{Cognitive input (pattern recognition/classification):} In this model, the data are created by professionals at a central facility, and volunteers provide cognitive input through distributed processing or classification. There are many popular CSR projects in this model with large public participation. Huge amounts of digitised data require a simple, well-defined task to be performed by online volunteers. Preliminary online training material familiarises participants with the identification/classification task. Images will appear in the online interface one after another to identify specific features. The identification tasks can range from identifying brightness dips in stellar light curves (used for the discovery of potential exoplanets) to recognising galaxy morphologies such as spiral, elliptical, barred spiral, or irregular. Participants have no control over the images being displayed; however, they can flag unusual or intriguing objects for discussion on project forums. To minimise human error in confusing cases or mistakes made by new participants, each image is independently classified by multiple participants before professional review. One prime example is Galaxy Zoo, launched in 2007 to classify galaxies from the images collected by the SDSS. This quickly became popular and has now been expanded to the Zooniverse project with dozens of science and humanities CSR projects.

One of the well-known success stories of Galaxy Zoo is the discovery of a relic ionised gas cloud near the spiral galaxy IC 2497, for which the source of ionising radiation, a quasar, switched off long before the epoch of observation. The CSR discoverer, Hanny van Arkel, a school teacher by profession, was awarded authorship in a mainstream research journal paper \citep{GalaxyZoo-Hanny-Lintott2009}. In that paper, with 20 authors, she earned the fourth author position for her contribution. Follow-up observations, including both deep optical imaging and long-slit spectroscopy, were performed to enhance the quality of this discovery paper. The newly discovered object was dubbed ``Hanny's Voorwerp'' (Dutch for `object'), and since then, dozens of new `voorwerps' have been spotted by Galaxy Zoo. Similarly, a few other new classes of objects discovered by Galaxy Zoo citizen scientists are Green Pea galaxies \citep{GalaxyZoo-GreenPeas-Cardamone2009}, red spirals \citep{GalaxyZoo-Red-Spiral-Masters2010}, and bulgeless AGNs \citep{GalaxyZoo-BulgelessAGN-Simmons2013}. 

Other successful examples of this model include Galaxy Cruize: galaxy morphology classification using deep optical images from the Subaru 8 m optical telescope \citep{CSR-GalaxyCruise-Tanak2023}, Planet Hunters: exoplanet detection using Kepler and Transiting Exoplanet Survey Satellite (TESS) light curves \citep{CSR-PlanetHunter-Schwamb2013}, Disk Detective: identification of debris disks around young stars using Wide-field Infrared Survey Explorer (WISE), Spitzer, and Hubble data \citep{CSR-DiskDetective-Kuchner2016}, Backyard World: Planet 9: search for faint moving objects such as brown dwarfs and Planet Nine candidates using WISE images \citep{CSR-BackyardWorld-Kuchner2017} etc. 

\textbf{Observational input (personal data collection):} In this model, the data are generated/collected by volunteers at distributed locations, while professionals perform analysis at the centralised locations. This form of CSR requires data to be collected at the time and location of the event, which is a basic requirement of the scientific investigation itself; a condition that centralised research facilities cannot fulfil. The role of the citizen scientist cannot even be replaced by an expert, as s/he can not be present in multiple locations at the same time. This has been particularly true in projects such as GLOBE at Night, a citizen science light pollution education project \citep{GLOBEatNight-Kyba2013NatSR...3.1835K}. Night sky luminescence measurement, air quality monitoring, and the environmental sensor network are a few examples. The iSPEX add-on to smartphones makes the quality of the data being collected uniform by enabling the same spectro-polarimeter instrument and reporting system to become almost omnipresent, simply because it can be attached to smartphones \citep{iSPEX-Burggraaff2020SPIE11389E..2KB}. A recent discovery of an atmospheric phenomenon known as Strong Thermal Emission Velocity Enhancement (STEVE) is also a beautiful example of this mode of CSR, where centralised processing of reports from multiple locations helps the progress of science that otherwise was not possible \citep{STEVE-MacDonald2018SciA....4...30M}.  Traditionally, observation of eclipses, meteors, and variable stars also falls in this class of CSR. The American Association of Variable Star Observers (AAVSO) is a prominent example of this model of CSR \footnote{\url{https://apps.aavso.org/jaavso/}}.

In summary, unlike in the previous model, the research instrument that generates the basic scientific data is in the hands of the citizen scientists, while the analysis of the collected data is conducted by experts at a central location in a research institute for quality international publications of a research paper or the publication of a database.  

\textbf{Computational input (no cognitive role):} This is primarily a passive mode of CSR where the participants do not contribute their human pattern recognition or intelligent interpretation of the data, but contribute their computational resources only when not otherwise in use. Search for Extraterrestrial Intelligence, or SETI@Home, is a well-known name in this model \citep{SETI-Anderson2025}. Since SETI is an important goal of human civilisation, numerous volunteers contribute their computer resources for the data analysis or signal processing of the data collected through the Arecibo radio telescope. This distributed computing approach to CSR, launched in 1999 by the University of California, became one of the earliest and largest examples of public involvement in scientific research. Although it was suspended in 2020, it has inspired many other CSR projects through distributed computing. Some project names are Einstein@home: Searches for gravitational waves and pulsars using data from LIGO and radio telescopes \citep{Einstein-Abbott2009}, Rosetta@home:  studies protein folding and design for biomedical research \citep{Rosetta-Kaufmann2010}, LHC@home: simulates particle accelerator experiments for CERN’s Large Hadron Collider \citep{LHC-Barranco2017}, etc. 

\textbf{Semi-professional input to Pro-Am Collaboratory:}  A professional–amateur collaboratory in which trained volunteers contribute interpretative insight and work closely with experts, supported by major research facilities to enhance the scientific output of a CSR project. This model represents a more advanced form of CSR, where trained volunteers work closely with professional astronomers in a collaborative framework. In this model, the initial discoveries made by the trained citizen scientists are scientifically promising but require deeper, higher-quality, or multi-wavelength follow-up observations that can only be obtained through major research facilities. This professional–amateur partnership allows the preliminary identification to be transformed into a publication-quality scientific result.

Research journals require submitted manuscripts to meet both quantity and quality standards for publication. It is possible that a newfound object from a CSR project is not good enough for direct publication and requires follow-up observation, either with deeper observation at the same wavelength or investigation in other wavelengths. Typically, CSR projects, by design, do not have guaranteed observatory times. Gems of Galaxy Zoos is a project in which follow-up observations of objects identified by Galaxy Zoo citizen scientists were carried out using the Hubble Space Telescope, which resulted in many beautiful publications \citep{ZooGEMS-Keel2022}.

Similarly, preliminary discovery reports submitted by the trained citizen scientists of the RAD@home Collaboratory are followed up with the GMRT for deeper/multi-frequency investigations. This observation, spread over multiple cycles of the Time Allocation Committee reviews, led to a better quality of publication than would have been possible without follow-up. For example, what was originally a wild interpretation, jet-galaxy interaction, based on NVSS, FIRST and TGSS images only, could be convincingly presented in a refereed publication thanks to new data from GMRT. This newfound object, named RAD-12, was found to be shooting a unidirectional radio jet which hits the neighbouring galaxy and bounces back, forming a mushroom bubble \citep{RAD12-Hota2022}.         
   
\section{CSR in Astronomy - A brief Chronology}
Some fields of science, such as biology, geology, and astronomy, have some aspects that require data collection or observations specific to a location on Earth; thus, these naturally lend themselves to CSR. Since ancient times, phenomena such as eclipses and meteors, which occur at specific locations and specific times, but are scientifically interesting to all, have been almost a natural form of CSR. For a broader understanding of public engagement in astronomy, through informal education systems institutionalised by planetaria and museums, including CSR, an extensive review is available in \cite{Pompea2020-ARAAA}. For a focused review on CSR in astronomy, we refer readers to \cite{Marshall-2015-ARAA}. However, for a simple understanding of how the field has evolved, a chronological history of CSR may be helpful.    

\textbf{Pre-1900:} Long before “citizen science” was coined, astronomy relied heavily on dedicated amateurs. Notable examples are Frederick William Herschel (discovered Uranus, 1781); Caroline Lucretia Herschel (discovered several comets); John Goodricke (discovered many variable stars, including the famous Algol); Antoine Darquier de Pellepoix (discovered a planetary nebula), etc. These amateurs often worked independently yet made fundamental discoveries because telescopes were accessible at an individual level and sky observations required patience and perseverance, rather than institutional support.

\textbf{Early 1900s:} The rise of organised amateur networks and systematic variable-star observing marked a major shift. Groups like the American Association of Variable Star Observers (AAVSO), founded in 1911, formalised the contributions of non-professional observers. Thousands of volunteers systematically monitored stellar brightness variations that remain scientifically valuable in professional studies today.

\textbf{Mid-1900s:} This was the era of photographic plate collections and amateur contributions. Observatories amassed millions of glass plates, and amateurs often helped to catalogue or identify transient events. Individuals such as Albert Francis Arthur Lofley Jones (New Zealand) and George Eric Deacon Alcock (UK) became famous for their comet and nova discoveries made visually or photographically.

\textbf{1980s–1990s:} Then came the era of the digital revolution and early computer-assisted searching. Affordable CCD cameras and personal computers transformed amateur astronomy. Amateurs began making scientifically calibrated measurements (photometry, spectroscopy). Networks like the British Astronomical Association Variable Star Section and the IAU Circulars began integrating amateurs into professional workflows for nova/supernova discoveries.

\textbf{1998–2005:} Then came the explosive era of internet-enabled distributed searches. The SETI@home project (1999) pioneered distributed computing, allowing volunteers’ personal computers to analyse radio signals for extraterrestrial intelligence. This marked the start of large-scale, data-driven public participation in astronomy.

\textbf{2007: } The internet-based participatory image analysis, or real CSR, was born. Galaxy Zoo was launched in 2007, inviting the public to classify galaxies from the SDSS. Within days, hundreds of thousands of volunteers participated, leading to major discoveries such as ``Hanny’s Voorwerp''. This success led to the creation of the Zooniverse platform, hosting dozens of astronomy (and non-astronomy) CSR projects.

\textbf{2010s:} In a few years, the CSR expanded into time-domain and multi-wavelength astronomy. Citizen scientists began contributing to supernova searches, exoplanet transits, and variable-star monitoring using both robotic telescope data and home observatories. Examples include Planet Hunters, Disk Detective, and Backyard Worlds: Planet 9. 

\textbf{2010–2020:} Without a custom-built web interface or using existing image analysis tools, various Professional-Amateur (Pro-Am) collaboratory models for CSR evolved. Near-Earth asteroid discovery from SDSS data led by the Spanish Virtual Observatory \citep{CSR-NEA-Solano2014}, and training and discovery of unusual radio sources in the TIFR GMRT Sky Survey (TGSS) by the RAD@home Astronomy Collaboratory \citep{RAD-Hota2014} are such examples. 

\textbf{2020s:} In recent times, we see the rise of AI-assisted, survey-scale CSR programmes. With the advent of huge datasets (LoTSS from LOFAR, MeerKAT, EMU from ASKAP), citizen scientists now help train AI models and verify rare objects found by algorithms. Hybrid approaches combine machine learning (ML) with human verification to handle terabytes of data. Interestingly, the results of cosmological simulations have also been compared with CSR using real observational data \citep{CSR-simulation-Dickinson2018}.

\vspace{-0.2cm}
\section{CSR in Radio Astronomy}
Radio astronomy covers a large range in distance to the objects it can investigate. It starts from understanding the local ionosphere, through galactic and extragalactic astronomy, out to the Epoch of Reionisation (EoR) and cosmic microwave background radiation (CMBR). Similarly, there are various science working groups in SKAO, namely Cosmology, Cradle of Life, Epoch of Reionisation, Extragalactic Continuum, Extragalactic Spectral Line, Gravitational Waves, High Energy Cosmic Particles, HI Galaxy Science, Magnetism, Our Galaxy, Pulsars, Solar, Heliospheric \& Ionospheric Physics, Transients and VLBI. Several CSR projects in radio astronomy have been listed on the SKAO website for the general public to get involved in and get oriented towards science with the SKA. Here, we describe a few CSR projects in radio astronomy that have proven popular using pre-SKA facilities (precursors and pathfinders) and may be useful to review before designing a CSR with the SKA.  

\vspace{-0.2cm}
\subsection{Radio Galaxy Zoo}
\begin{figure}[h]
    \centering
	\includegraphics[height=0.3\columnwidth]{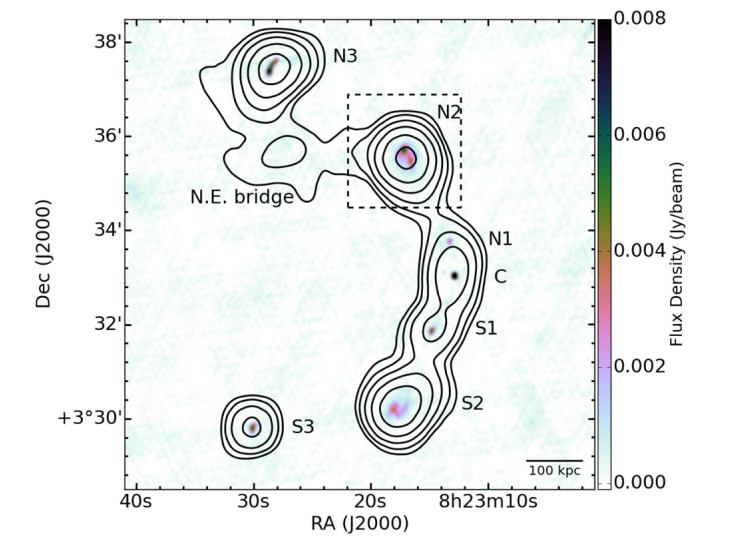}
    \includegraphics[height=0.3\columnwidth]{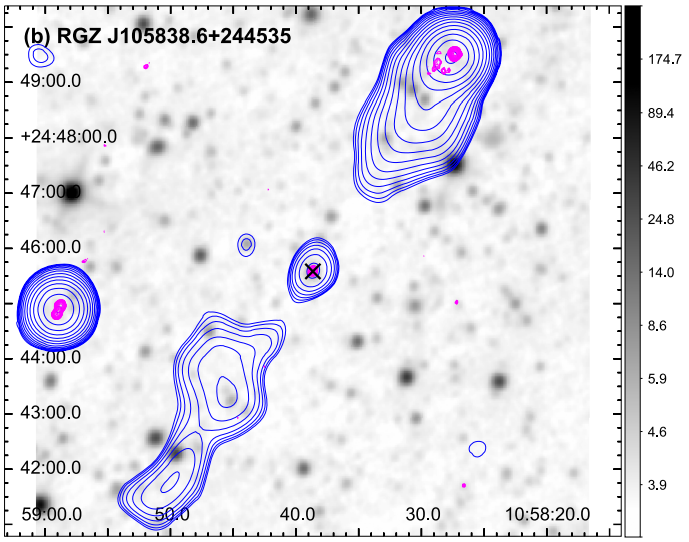}
            \caption{Left panel: Image of the episodic wide-angle tailed radio galaxy (RGZ J082312.9$+$033301) discovered by the RGZ project from \cite{RGZ-Banfield2016MNRAS}. Right Panel: A HyMoRS from the same RGZ project displaying FR-I nature on the southern and clear edge-brightened FR-II nature on the northern side from \cite{RGZ-HyMoRS-Kapinska2017}. }
    \label{fig:fig2}
\end{figure}
Radio Galaxy Zoo (RGZ) is a CSR project launched in 2013 on the Zooniverse platform to involve the public in classifying and identifying radio galaxies \citep{RGZ-Banfield2015}. The project addresses a major challenge in radio astronomy: modern radio surveys detect millions of sources, but associating radio emissions with their host galaxies in optical or infrared images requires human pattern recognition that computers cannot yet reliably perform. In RGZ, volunteers examine images that overlay radio data (from surveys like FIRST and ATLAS) on infrared backgrounds (from WISE and SWIRE). Participants identify which radio components belong to the same galaxy and mark the likely position of the “host galaxy”. Each image is independently classified by many users, and their collective results are aggregated to produce reliable catalogues of radio galaxies.
The project has attracted tens of thousands of volunteers worldwide and has yielded significant scientific results. RGZ citizen scientists have helped discover giant radio galaxies \citep[GRGs;][]{RGZ-Giants-Tang2020}, hybrid morphology radio sources \citep[HyMoRS;][]{RGZ-HyMoRS-Kapinska2017} with peculiar lobe structures that expand our understanding of galaxy evolution and black hole activity (Fig.~\ref{fig:fig2} right panel). Results from RGZ have been published in peer-reviewed journals, more than 15 so far, and some volunteers, with significant contributions, have been recognised as co-authors.

A good example is the discovery of RGZ J082312.9$+$033301, a giant episodic Wide Angle Tailed (WAT) radio galaxy (Fig.~\ref{fig:fig2} left panel), suggesting a poor-cluster environment \citep{RGZ-Banfield2016MNRAS}. The published research paper awards authorship to the discoverer citizen scientists T. Matorny and I.A. Terentev, with the Zooniverse project as their affiliation. Interestingly, after the initial finding of the object, so much further analysis of the archival data has gone into it that the author list contains eighteen authors from five different countries, and the citizen scientists are placed ahead of other professional astronomers. The discovery of the object, added with multi-wavelength archival data, was significant enough that it did not require further follow-up observation before publication. Beyond its discoveries, RGZ serves as a training ground for developing ML algorithms, using human classifications to teach computers how to recognise radio galaxy patterns. The project also acts as a model for future large-scale radio surveys, including those from ASKAP, MeerKAT, and the upcoming Square Kilometre Array (SKA). By combining the power of public participation with professional research, RGZ has demonstrated that citizen scientists can make genuine contributions to frontier astrophysics, helping to map and understand the radio universe. 

\subsection{Radio Galaxy Zoo: LOFAR}
The Radio Galaxy Zoo: LOFAR (RGZ: LOFAR) project is a large CSR effort that pairs human visual classification with algorithmic cross-matching to identify radio sources and their optical/IR host galaxies in wide-area low-frequency radio surveys conducted with the LOw Frequency ARray (LOFAR). RGZ: LOFAR extends the RGZ concept to the unprecedented surface-brightness sensitivity and high angular fidelity of LOFAR Two-Metre Sky Survey (LoTSS) images, enabling volunteers to inspect complex, extended radio structures that automatic source-finders frequently fragment or disassociate. The primary scientific goals are to produce robust source associations and radio–host identifications, classify radio morphologies (e.g., compact core, jets, lobes, double, bent/tailed), and thereby facilitate population studies of active galactic nuclei (AGN), star-forming galaxies and rare extended systems (GRGs, restarted/episodic sources, remnant lobes). 

LoTSS provides the foundational data for the project: it generates wide-field 120–168 MHz maps with arcsecond-level, direction-dependent calibrated images that reveal diffuse, low-surface-brightness emission on large scales. Because LoTSS detects a very large number of resolved and multi-component sources, automated cataloguing and cross-identification (e.g., source finders and likelihood-ratio methods) must be supplemented by visual association to avoid incorrect host assignments and to capture morphological complexity. RGZ: LOFAR therefore integrates volunteer classifications with ML algorithmic approaches developed by the LoTSS team \citep[e.g.,][]{Barkus+22}; the outputs feed into LoTSS data releases and downstream science catalogues, improving both completeness and reliability of optical identifications. 

Operationally, RGZ: LOFAR runs on the Zooniverse platform and presents volunteers with radio cutouts (normally overlaid on optical/IR images) together with simple, guided questions to indicate the number of radio components belonging to a single physical source and to mark the most likely host galaxy \citep{RGZ-LOFAR-Hardcastle2023}. The project workflow captures multiple independent classifications per object; consensus algorithms and expert vetting are then used to produce a high-quality association catalogue. This human-in-the-loop procedure is particularly valuable for extended or asymmetric radio sources, cases where ML algorithms either fragment emission into separate entries or fail to link jets and lobes to a central host. The project proved popular: initial runs amassed large numbers of volunteer classifications in weeks, enabling a rapid, community-driven inspection of bright and complex LOFAR sources. 

Scientifically, RGZ: LOFAR serves several interconnected purposes. First, it directly supplies visual classifications and host associations used to construct LoTSS source catalogues and to validate automated cross-matching pipelines (likelihood-ratio, ML classifiers). Second, the labelled sample creates training and test datasets that are indispensable for developing ML models tailored to low-frequency radio morphology and association tasks \citep[e.g.,][]{Alegre+24}; such models aim to scale classification to the millions of sources LOFAR and future SKA-pathfinder surveys will detect. Third, the human classifications enable targeted studies of rare populations (restarted radio galaxies, double-double systems, GRGs, bent-tail sources), environmental influences on radio morphology, and the demographics of radio AGN across cosmic time. Recent LoTSS data releases explicitly combine ML algorithmic identification with RGZ: LOFAR results to produce the most complete optical identifications to date for the northern LOFAR sky \citep{RGZ-LOFAR-Hardcastle2023}. This paper, making use of nearly one million CSR classifications from over 10,000 volunteers, awarded authorship to all the most active CSR participants who agreed to sign up to the publication, 12 individuals in total. \citet{Horton2025} used LoTSS DR2 together with LOFAR Galaxy Zoo visual classifications to assemble a sample of 9985 extended radio galaxies, of which 7613 have reliable redshifts. Morphological inspection yielded 6231 FRII, 2922 FRI, 966 hybrid, 234 spiral, and 440 relaxed-double systems. The work highlights the reliability and scientific value of Galaxy Zoo-based classifications for large LOFAR samples.



It is a critical bridge linking LOFAR’s rich, low-frequency imaging to robust science products. Its human classifications correct, validate and train automated methods; enable discovery and statistical studies of complex radio source populations; and provide a scalable template for CSR integration in future large surveys. As LoTSS continues to expand and as ML models mature using RGZ: LOFAR training sets, the combined human-and-AI strategy will be central to exploiting the SKA-era radio sky.

\vspace{-0.2cm}
\subsection{RAD@home Astronomy Collaboratory}
{\bf A scalable CSR model advancing extragalactic radio astronomy:}
Since its foundation on 15 April 2013, RAD@home\footnote{\url{https://radathomeindia.org/}} has established itself as India’s first CSR platform in astronomy, with the mission of enabling any motivated undergraduate (BSc/BE) or graduate student to participate in multi-wavelength extragalactic radio astronomy research from home. It operates as a “zero-funding, zero-infrastructure” national model, functioning entirely through online groups and experiential training that connects volunteers with professional astronomers and large-survey data resources. The platform follows an “Online Identification to Online Publication” workflow: trained citizen scientists (referred to as e-/i-astronomers) identify candidate radio sources, professional collaborators conduct follow-up analysis (archival or observational), and peer-reviewed papers are co-authored with the participating e-astronomers \citep{RAD-Hota2014, RAD-Hota2016}.

The model prioritises scalability and inclusivity. By leveraging free archival survey data and internet-based collaborative sessions, RAD@home enables research participation across socio-economic backgrounds and independent of institutional affiliation, opening pathways into frontline research for individuals who would otherwise be excluded from traditional academic environments.

{\bf Training and operational model:}
RAD@home employs a structured, step-wise training workflow to build citizen-scientist capability in radio continuum and multi-wavelength extragalactic astronomy. At the entry level, participants join through a Facebook group, where they learn to analyse multi-wavelength Red-Green-Blue (RGB)$+$contour composite images of galaxies using UV, optical, infrared, and radio data. With a simple online tool (RAD-RGB-maker)\footnote{\url{https://radathomeindia.org/rgbmaker}}, participants generate RGB images in a standard format and post them in the discussion group for activity-based learning \citep{RAD-Kumar2023}. Each day, a different galaxy—typically from the NGC or 3C catalogues is discussed, forming the activity known as ``Daily Galaxy RGB Analysis''.
For example, on a given day, participants analyse galaxies NGC 815, NGC 1815, NGC 2815, NGC 3815 up to NGC 7815. 
Although RGB images may be posted at any time, discussions peak during a designated session each Saturday afternoon (2–3 pm), typically generating a high volume of image posts and analytical comments.

Once a participant demonstrates proficiency, referred to as becoming “RGB-qualified”, they are promoted to the i-astronomer level, where they are trained to analyse FITS images from the GMRT TGSS survey. In addition to this online entry route, RAD@home also runs an offline pathway. One-Day RAD@home Astronomy Workshops (ODRAW) are conducted free of cost at educational institutions across India. Instead of Daily Galaxy Analysis, participants engage in the “Nationwide Inter-University RGB-Contouring analysis” (NIU-galaxy RGBC analysis), where the last three digits of a participant’s postal (PIN) code determine the galaxy to analyse. If the digits are abc, the galaxy becomes NGC xabc, where x ranges from 0 to 7. Participants create RGB+contour images using the RAD-RGB-maker and discuss them during the workshop, gaining practical multi-wavelength image-analysis experience. From here, participants may be invited to join the i-astronomer group for regular online training and research discovery.

Through lectures and guided discussions, i-astronomers are introduced to concepts including galaxy mergers, AGNs, radio galaxies, and galaxy–black hole co-evolution. They are trained to recognise morphological classes such as FR-I, FR-II, wide-angle tail, head-tail, X-shaped, and double–double radio galaxies. The key objective is to detect sources that deviate from these standard classes—objects that appear unusual, faint, or morphologically ambiguous. Using combined NVSS–TGSS–FIRST radio data and DSS optical overlays generated from the RGB-maker tool, trained participants begin identifying candidate discoveries, which are checked against the literature.

The advanced stage of training occurs through RAD@home Discovery Camps, offered both offline and online. RGB-qualified participants from across India attend a week-long residential camp — hosted free of cost by partnering institutions -- or the online camp version held on weekends over a month. Upon successful completion, they are promoted to the e-astronomer group, where active research takes place.

Research discussions with e-astronomers are held in weekly online e-classes every Saturday from 3–4 pm. Potential discoveries are posted in the group without revealing the exact coordinates (RA/Dec), enabling collaborative analysis while protecting the discoverer’s ownership of the object. When the group collectively agrees that a candidate appears scientifically significant, the discoverer submits a First Investigation Report (FIR) via a formal Google Form \citep[for more details, refer to ][]{RAD-Hota2016}. The object is then evaluated by professional astronomers (PhD-level members of the collaboratory), who determine whether it should proceed directly to publication or if follow-up observations are required. When further observations are needed, proposals are submitted under the programme “GMRT Observation of Objects Discovered by RAD@home Astronomy Collaboratory (GOOD-RAC).” 

{\bf Scientific contributions:} 
\begin{figure}[h]
    \centering
	\includegraphics[height=0.20\columnwidth]{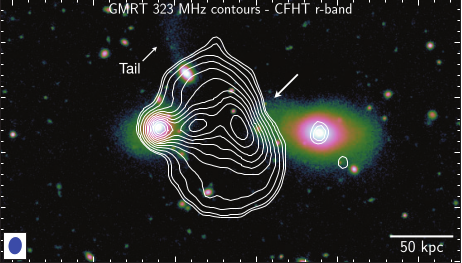}
     \includegraphics[height=0.20\columnwidth]{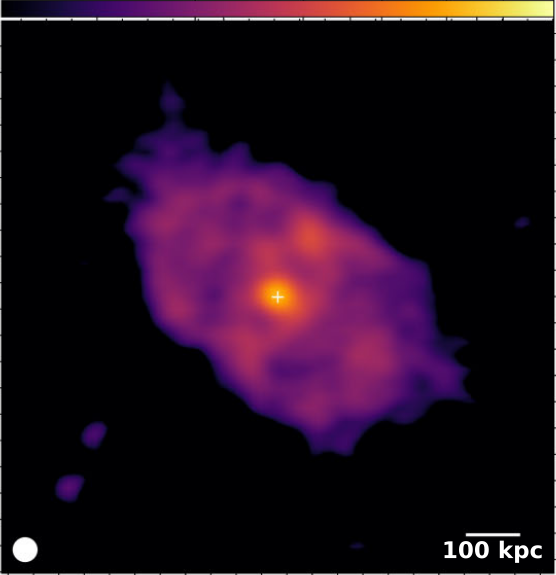}
     \includegraphics[height=0.20\columnwidth]{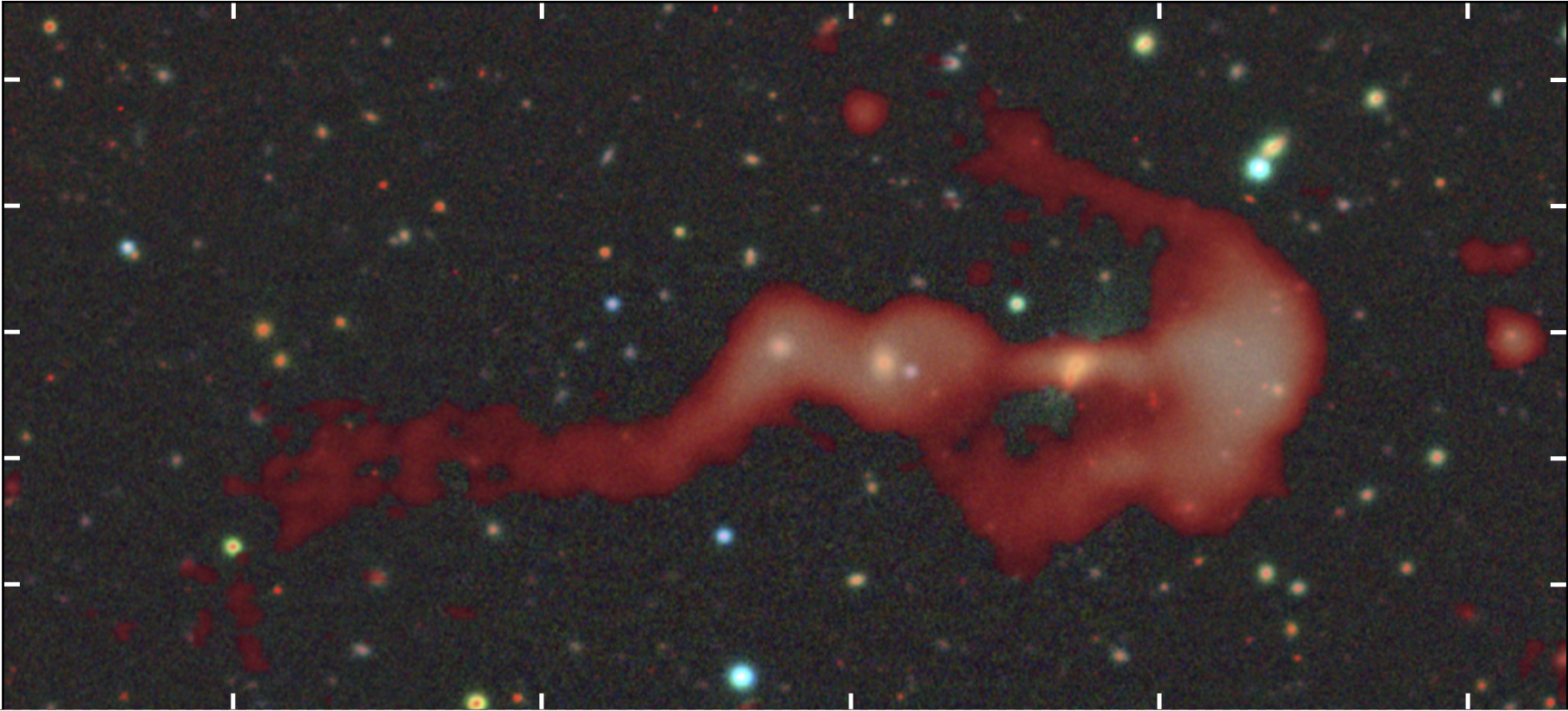}
           \caption{Three examples of RAD@home CSR discoveries. Left Panel: A unique case of jet-galaxy interaction in RAD12 from \cite[]{RAD12-Hota2022}. 
           Middle Panel: RAD J131346.9$+$500320, the farthest and most powerful Odd Radio Circle discovered by the RAD@home collaboratory from \cite{RAD-ORC-Hota2025MNRAS}. Notice the twin intersecting rings embedded inside the extended radio emission. 
           Right panel: A recently discovered unique Bow-and-Arrow-shaped Radio Galaxy (RAD-BAARG) possibly tracing a large-scale ($\sim$560 kpc) leading-edge shock as the galaxy falls into a nearby cluster environment from \cite{RAD-BAARG-Hota2026MNRASL}. }
    \label{fig:fig3}
\end{figure}

Over the past decade, RAD@home has produced several peer-reviewed research papers and notable discoveries in the field of extragalactic radio astronomy. The primary scientific goal of the collaboratory is to investigate AGN feedback and merger-driven galaxy evolution using a multi-wavelength approach, with a particular emphasis on low-frequency radio observations. For example, \cite{RAD12-Hota2022} reported a CSR-driven discovery of an AGN generating a large unipolar radio bubble ($\sim$137 kpc) that appears to interact with its merging companion galaxy --- an extremely rare phenomenon that provides a unique laboratory to study jet-galaxy interactions (refer to the left panel of Fig.\,\ref{fig:fig3}). Similarly, a rare Collimated Synchrotron Thread (CST; RAD J083109.5$+$414738) connecting the eastern and western radio lobes has been identified by \cite{RAD-Hota2024}. The nature and origin of CST are still a mystery, and SKAO can help resolve it (see review in this volume by \cite{Fromm01.2026.SKA}). Recently, \cite{RAD-Apoorva2025} identified two spiral-host radio galaxies through RAD@home in which the radio lobes extend over 30-220 kpc and display asymmetries consistent with ram-pressure stripping of the host galaxy discs. In addition, RAD@home has published a catalogue of ten unusual radio sources, including a candidate jet-galaxy interaction system (RAD-Q-mark galaxy; RAD J091859.3$+$315140), a giant episodic radio galaxy (RAD-giant DDRG; RAD J111142.6$-$132411), remnant/relic radio galaxies (RAD-Rabbit; RAD J121934.9$-$132611), and a giant extremely asymmetric episodic radio galaxy (RAD  J121911.5$+$610816), among others \citep[for more details refer to ][]{RAD-Hota2024}. For a review on the study of such episodic radio galaxies helping to understand AGN duty cycle and giant radio galaxies helping to trace large-scale structures like galaxy filaments and clusters, two reviews by \cite{Hardcastle01.2026.SKA} and \cite{Dabhade01.2026.SKA}, respectively, can be seen in this volume.

The most striking achievement to date is the publication of three rare radio sources discovered by RAD@home volunteers. One of these is the first "Odd radio circle" (ORC) discovered through a CSR process: RAD J131346.9+500320 ($z\sim 0.94$). ORCs are a recent puzzle in radio astronomy. Since its discovery in 2021, multiple models for its origin have been proposed, but none can satisfactorily explain all the ORCs  \citep{ORC-discovery-Norris2021, ORC-physics-Norris2022}. This newly discovered ORC, RAD J131346.9+500320, exhibits two intersecting rings, each $\sim$300 kpc in diameter, and is embedded within an extended diffuse halo $\sim$800 kpc across (refer to the middle panel of Fig.\,\ref{fig:fig3}). The other two sources are GRGs with ring-like structures at the termini of their jets, both situated within massive cluster-scale haloes $\sim 10^{14}$\msun \citep{RAD-ORC-Hota2025MNRAS}. The RAD@home ORC is currently the farthest and the second-most powerful ORC known, and importantly, it represents the first ORC detected using low-frequency LOFAR data. A recent publication containing a list of new ORCs discovered using LOFAR can be seen in \cite{ORC-LOFAR-DeGasperin2026}. For a review on ring-like circularly symmetric structures, ORCs and circular radio relics, articles by \cite{SabyasachiPal02.2026.SKA} and \cite{Koribalski01.2026.SKA} can be referred to in this volume.

Very recently, RAD@home discovered a unique Bow-And-Arrow-shaped Radio Galaxy (RAD-BAARG) that traces a large-scale ($\sim$560 kpc) bow-shock ahead of an elliptical galaxy (refer to the right panel of Fig.\,\ref{fig:fig3}). It is likely the first radio band detection of a leading-edge shock as a galaxy falls into a nearby cluster-scale environment \citep{RAD-BAARG-Hota2026MNRASL}. The radio(red)-optical(RGB) image also shows the large radio tail on the other side with S-shaped distortion, which is a possible sign of either jet-galaxy interaction or instability. Future sensitive images with the SKAO may reveal many such dynamical processes in the circum-galactic medium or cosmic web, revealed by radio-plasma supplied by large-scale AGN jets. See a review by \cite{Dabhade01.2026.SKA} in this volume on giant radio galaxies as a tracer of large-scale environment.  

These discoveries highlight the unique value of human visual pattern recognition in identifying rare and complex radio morphologies that automated pipelines frequently overlook. They also demonstrate the emergence of a novel CSR-enabled discovery pipeline capable of operating effectively in the era of massive radio surveys, and suggest that RAD@home is well positioned for scientific contributions in the upcoming SKA era.

{\bf Impact on research culture and capacity building:}
Beyond producing refereed publications, RAD@home’s greater significance lies in its contribution to human-resource development and the democratisation of research participation. It shows that citizen science can function as a genuine research platform, generating citable scientific results with trained volunteers included as co-authors. By enabling non-institutional participants to analyse real multi-wavelength survey data, RAD@home broadens the scientific talent pool, particularly for students and learners outside major research centres.

Because it operates fully online and relies on publicly accessible archives, RAD@home offers a scalable, low-cost model for resource-limited regions to engage with frontier astronomy. Its “ABCDresearch” philosophy (Any BSc/BE Can Do research) is reflected in a structured training pipeline that prepares participants for next-generation facilities such as the SKAO.  For details on this socio-scientific process that has evolved over the years and aligning human resource development towards participation in SKAO, readers can refer to \cite{RAD-Hota2016}. After more than a decade of refinement, RAD@home is recognised as a framework for research-grade capacity building \citep{RAD-Bagla2025, RAD-Dewangan2025, RAD-Kale2025}, enabling volunteer co-discovery of rare radio phenomena/features while cultivating a new generation of citizen astronomers.

\subsection{Brief notes on other CSR in radio astronomy}
{\bf Radio Galaxy Zoo: EMU:} The project helps astronomers locate and identify supermassive black holes, star-forming galaxies, and maybe discover something completely unknown with an SKA pathfinder telescope, the Australian Square Kilometre Array Pathfinder (ASKAP). Following the original, very productive RGZ, the team behind it is asking the public to analyse new radio images from the Evolutionary Map of the Universe (EMU) survey from the ASKAP \citep{EMU-Norris2021, EMU-Hopkins2025}. 
Participants are asked to identify fragments of a radio source as a unit, together with identifying the host galaxy from optical/IR images. The project offers a chance to take on the challenging task of describing what each radio source looks like, using a preselected list of descriptive tags that have been defined using machine learning methodologies such as anomaly detection and natural language processing \citep{RGZ-EMU-Bowles2023}. Due to its higher frequency (944 MHz), better diffuse emission sensitivity (25 $\mu$Jy beam$^{-1}$) and reasonably high angular resolution($\sim$ 13$^{\prime\prime}$), the survey is not only revealing new radio sources in the sky but also revealing new faint and diffuse structures in known radio sources. These sources, from a pilot survey, have been visually classified and will be used to train the ML algorithms and ultimately classify sources automatically when a significantly large volume of survey data becomes available \citep{EMU-DRAGN-Norris2025}. 

{\bf Bursts from Space: MeerKAT:} 
Unlike traditional radio continuum sky surveys such as NVSS, FIRST, and TGSS, which were primarily designed to produce static maps of the radio sky, the exploration of the dynamic radio universe requires a fundamentally different observing strategy. Many astrophysical sources exhibit variability on timescales ranging from seconds to years: some undergo periodic changes in brightness, while others produce rare, one-off explosive events such as bursts and flares. To capture such transient phenomena, the MeerKAT radio telescope conducts repeated observations of the same regions of the sky, often revisiting fields on a weekly basis. This observing cadence enables astronomers to identify and monitor radio sources whose brightness changes over time, opening a new window on the variable and transient radio sky.

The vast quantity of data generated by these repeated observations presents a significant challenge for conventional analysis methods. To address this challenge, the Bursts from Space: MeerKAT citizen science project was launched on the Zooniverse platform, allowing members of the public to participate directly in frontline astronomical research \citep{Burst-MeerKAT-Andersson2023}. Citizen scientists inspect radio images and light curves in search of transient and variable sources that may be associated with a wide range of astrophysical phenomena, including stellar flares, supernovae, pulsars, variable AGNs, OH maser stars, X-ray binaries, and relativistic jets. Remarkably, more than a thousand volunteers contributed classifications, collectively examining thousands of candidate sources in only a few months. Their efforts led to the identification of 168 radio transient and variable sources, producing one of the largest catalogues of candidate radio variables assembled to date. Citizen scientists successfully recovered many known transients, uncovered previously unrecognised variable sources, and demonstrated an impressive ability to recognise complex patterns in noisy radio data that remain challenging for automated algorithms. Analysis of the resulting catalogue revealed that nearly 2\% of AGNs exhibit measurable radio variability, while also highlighting rare and unusual objects worthy of detailed follow-up observations. These results demonstrate that citizen science is not merely an outreach activity but a powerful scientific methodology that enables the efficient exploration of the time-variable radio sky and provides valuable training datasets for the development of next-generation machine-learning tools needed for future facilities such as the SKAO.

{\bf Pulsar Search Collaboratory (PSC):}
PSC trains students and citizens to analyse pulsar survey data and identify new pulsars. It uses the data collected from the Robert C. Byrd Green Bank Telescope for the purpose of discovering new pulsars. PSC has discovered several pulsars, including some millisecond pulsars \citep{CSR-PulsarSC-Rosen2013}.

{\bf PULSE@Parkes:} PULSE@Parkes is a free educational program where high school students use the CSIRO Parkes radio telescope, Murriyang, live and remotely to observe pulsars, analyse their data and interact with professional astronomers \citep{PULSE-Hobbs2009}.

{\bf Einstein@Home:} Einstein@Home uses the idle time of participating computers to search for weak signals from spinning neutron stars (or pulsars) using data from the LIGO gravitational-wave detectors, archival data collected in the past with the Arecibo radio telescope, MeerKAT radio telescope and the Fermi gamma-ray satellite. By joining, participants help researchers find more pulsars, indirectly! Pulsars are extremely interesting objects that radio telescopes like the SKA will use to observe gravitational waves \citep{Einstein-Abbott2009}. 

{\bf SETI@home:} SETI@Home uses distributed computing to search for extraterrestrial radio signals from the Arecibo Telescope data \citep{SETI-Anderson2025}. The project pioneered large-scale distributed citizen computing in radio astronomy.  Over 5 million volunteers participated in it before the project hibernated in 2020.

{\bf Radio JOVE:} The NASA-sponsored Radio JOVE project is a hands-on, inquiry-based educational project that allows students, teachers and the general public to learn about radio astronomy by building their own radio telescope from an inexpensive kit (155 USD + shipping in the year 2006) to observe Jupiter, the Sun and even our own galaxy at  20.1 MHz \citep{RadioJOVE-Thieman2006}.

{\bf RADIIO:} Finally, we describe another project, like RAD@home, that not only pursues scientific discovery through citizen science but focus more on development through astronomy. Radio Astronomy Development, Innovation and Inclusion for Outreach (RADIIO) is an international educational and capacity-building initiative funded by the International Astronomical Union Office of Astronomy for Development (IAU-OAD). The project uses radio astronomy as a tool for education, skill development, and scientific inclusion, particularly in regions with limited access to advanced research infrastructure. It helps participants build small, low-cost radio telescopes to detect the neutral atomic hydrogen line (21cm) from the plane of our Milkyway galaxy. RADIIO aims to make radio astronomy more accessible to students, teachers, and citizen scientists by developing hands-on learning resources, training programs, and collaborative research opportunities. The project emphasises interdisciplinary learning by combining astronomy with physics, signal processing, data science, electronics, and computational thinking. A central goal of RADIIO is to foster global participation in astronomy by building networks between researchers, educators, and learners across different countries. They participate directly in the Radio Galaxy Zoo-EMU project and become part of discoveries in radio astronomy. From making low-cost radio telescopes to discovering with large, world-class telescopes provide the participants gain a complete experience of personal growth through astronomy. Beyond teaching scientific concepts, RADIIO promotes inclusive participation in authentic research, enabling learners to contribute meaningfully to data analysis and discovery and, prepare them for the era of SKAO. In this way, RADIIO supports the broader mission of the IAU OAD: using astronomy to advance education and sustainable development worldwide.

\section{Synergy of CSR with AI/ML}
\begin{figure}[h]
    \centering
	\includegraphics[height=0.3\columnwidth]{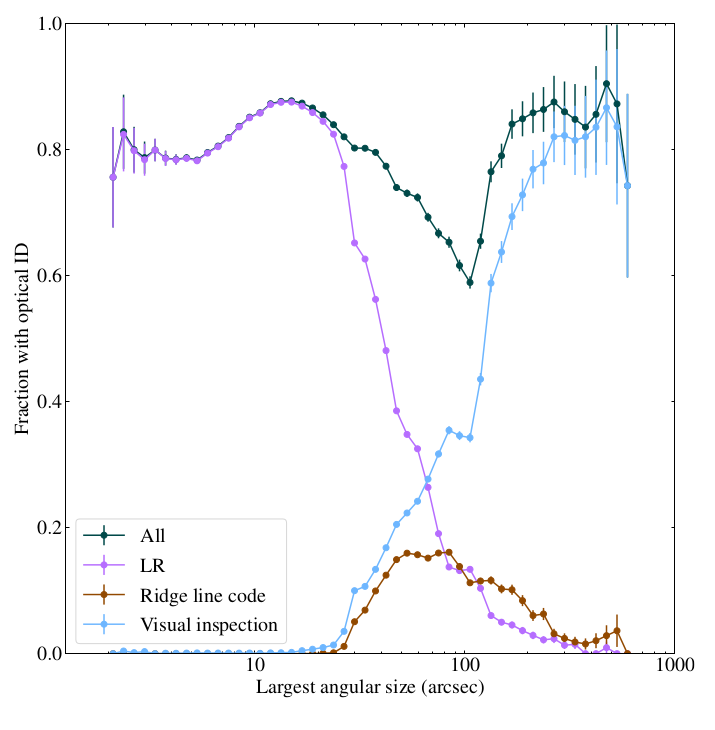}
    \includegraphics[height=0.3\columnwidth]{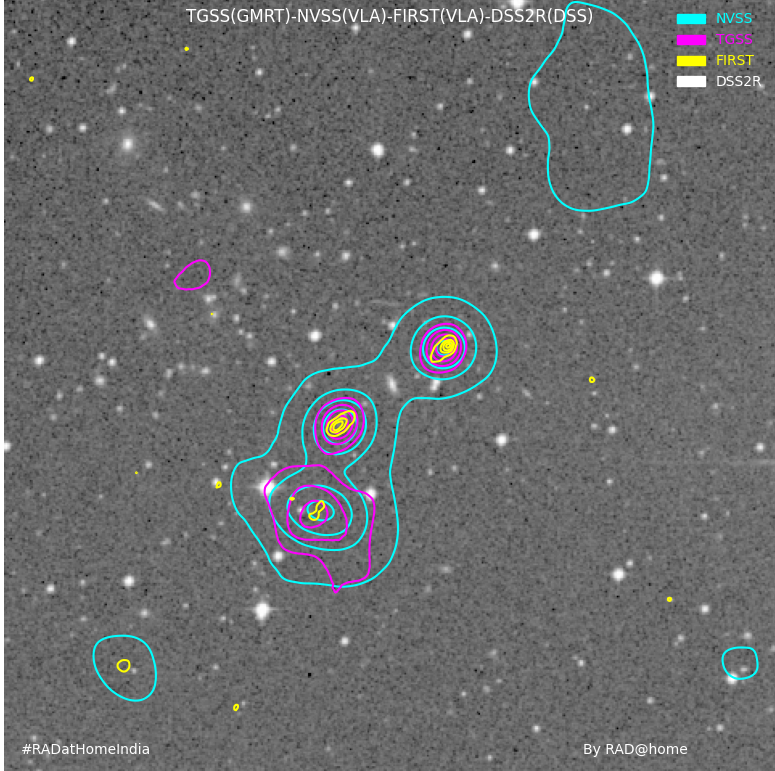}
    \includegraphics[height=0.3\columnwidth]{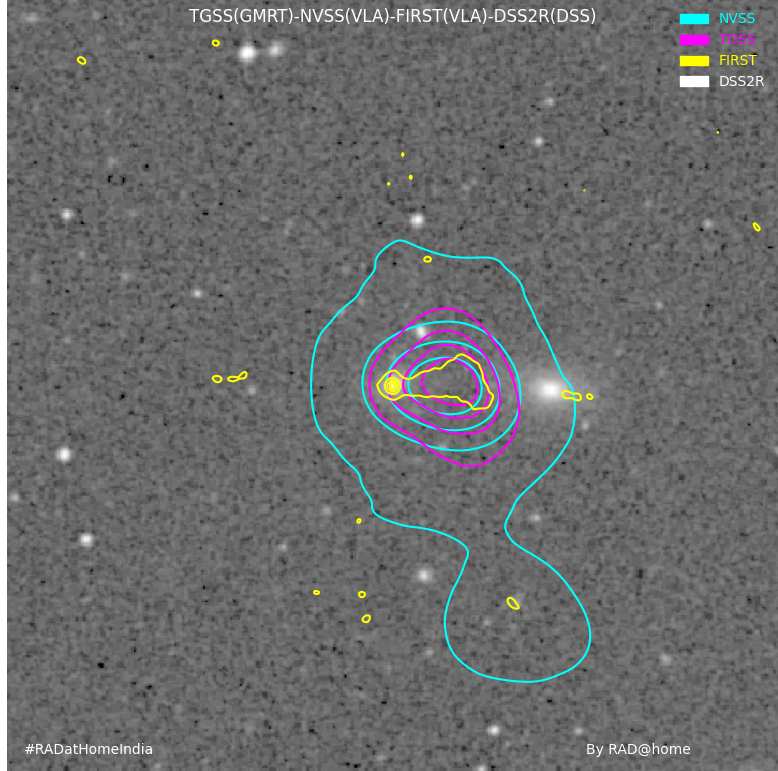}
        \caption{Left Panel: Comparison of visual classification by RGZ: LOFAR citizen scientists and expert human inspection with source classification by automated algorithms from \cite{RGZ-LOFAR-Hardcastle2023}. Visual inspection dominates the source identifications for the large, complex sources. Middle panel: As an example, episodic GRG `Speca' from NVSS (cyan), TGSS (magenta) and FIRST (yellow) are plotted in contours on DSS optical \citep{Speca-Hota2011}. This complexity requires human imagination to figure out that it is a single radio galaxy. Right Panel: With similar contours and grey scale scheme RAD12, case for jet-galaxy interaction, is presented \citep{RAD12-Hota2022} .} 
    \label{fig:example_figure}
\end{figure}
The collaboration between CSR and artificial intelligence (AI) is reshaping the landscape of modern research by merging human curiosity and creativity with machine intelligence and forming a unified system of discovery.  CSR, which is the active involvement of non-professionals in the scientific research process, provides human insight, contextual understanding, and large-scale participation. AI, on the other hand, offers computational power, speed, and scale in pattern recognition and automation capabilities. Together, they create a synergistic framework that enhances both scientific productivity and public engagement. This new human-machine collaboration is almost reinventing how we make discoveries in the future.

AI and ML have become indispensable for analysing massive datasets generated by modern telescopes and catalogue outputs of CSR. However, the reliability of AI models depends on the quality and diversity of their training data. Citizen scientists contribute significantly by producing, labelling, and validating large datasets that feed AI algorithms. For example, the Galaxy Zoo project \citep{GalaxyZoo-Lintott2008-paper1} mobilised 
over $\sim$10$^5$ volunteers to perform  $\sim$4 $\times$ 10$^7$ individual classifications of galaxies from the SDSS. The dataset generated from here was used later on to train ML systems capable of autonomous galaxy classification. A ``gold'' sample of high-agreement votes was fed as the labelled training data to a neural network classifier that efficiently reproduced the morphological classes (spiral, elliptical, and imaging artefact) of SDSS galaxy images \citep{GalaxyZoo-ML-Banerji2010MNRAS.406..342B}. Computer-generated visual morphological classifications of 3 million galaxies were produced, where Galaxy Zoo human classifications were used for training and validation \citep{AI-ML-SDSS-Kuminski2016ApJS..223...20K}. 
In some recent works, the EMU-RGZ programme tried to combine CSR and AI to accurately classify complex radio sources and create high-quality open-science catalogues named EMUCAT
\citep{EMU-AI-Vardoulaki2025arXiv250919787V, EMU-Tang2025}. This collaboration demonstrated how human classification efforts could bootstrap AI capabilities, establishing a foundation for data-driven astronomy. With time, AI technology is advancing at such a rate that the machine can overtake pattern recognition or anomaly detection in data \citep{AI-EMU-Segal2023}. Thus, promoting a collaboration between human and machine will enable more robust and faster results and human development, which is essential in the data-intensive and mega-telescope era we have already entered.

However, such human-machine collaboration must be guided by transparency, data ethics, and inclusivity \citep{CSR-Irwin1995_CitizenScience, RasmussenCooper2019_CitizenScienceEthics}. 
Citizen participation should complement institutional science rather than be exploited for low-cost labour. Responsible integration of AI with CSR must ensure mutual learning, acknowledgement, and equitable access to technological benefits.

It is important to recognise that AI can only identify the specific patterns for which it has been trained. Although it can rapidly classify large datasets and generate extensive catalogues, scientific discovery inherently lies beyond predefined categories. Discovery requires the ability to recognise the unfamiliar, interpret anomalies, and anticipate phenomena that have never been seen before. As the saying goes, discovery favours the prepared mind. It is not the structure or colour of the image that makes a discovery, but the imagination about the new phenomena it represents. Hence, many times the eye-brain system will outperform ML in making true discoveries.

A clear example, already discussed above, is the identification of the giant Wide-Angle Tail (WAT) radio source RGZ J082312.9$+$033301 by volunteers in RGZ. The object is neither a standard WAT nor a typical episodic or double-double radio galaxy (DDRG), making its classification difficult for an automated system trained on conventional morphology \citep{RGZ-Banfield2016MNRAS}. Similarly, RAD12—an asymmetric, one-sided jet impacting a neighbouring galaxy—did not match standard algorithmic expectations for an FR-II radio galaxy. The presence of two optical galaxies flanking the radio peak prompted human volunteers to recognise that something unusual was occurring, leading to targeted follow-up observations with the GMRT and then a publication \citep{RAD12-Hota2022}.

Another striking case where human intuition surpassed automated algorithms is the discovery of the farthest and most powerful ORC, RAD J131346.9$+$500320, by the RAD@home community \citep{RAD-ORC-Hota2025MNRAS}. Earlier automated pipelines misclassified this source as a simple elongated double because the ML model had been trained specifically to search for GRGs \citep{RAD-ORC-ML-giant-Mostert2024}. The algorithm had no conceptual framework for recognising an ORC, and the faint twin ring surrounded by diffuse emission structure fell outside its classification space. In other words, AI could not discover what it was never taught to expect.

At the scale of large samples, citizen scientists also outperform algorithms in characterising extended radio morphology. \cite{RGZ-LOFAR-Hardcastle2023} demonstrates (see Fig.\,\ref{fig:example_figure}, left panel) that RGZ volunteers are more effective than automated methods in the identification and association of multi-component radio sources in LOFAR images. Humans are uniquely capable of noticing anomalies and synthesising contextual information from multiple datasets. What appears as a faint, barely perceptible blob in the FIRST survey may be significantly brighter in NVSS, suggesting the presence of a relic lobe associated with an otherwise standard radio galaxy (Fig.\,\ref{fig:example_figure}, middle panel), as in the case of giant DDRG, Speca \citep{Speca-Hota2011}. Citizen scientists can be trained to routinely utilise powerful multi-wavelength exploration platforms such as Aladin Sky Atlas \citep{Bonnarel2000}, combining radio, optical, UV, infrared and X-ray information to pursue unexpected leads and deepen the investigation.

Thus, while AI accelerates data handling and routine classifications, citizen scientists excel at recognising the unusual, making intuitive connections, and driving discovery into unexplored parameter spaces. The synergy between AI and human pattern recognition is not only beneficial but essential to progress in modern astronomy \citep{Fortson2024_AI_CitizenScience}.

\vspace{-0.2cm}
\section{Citizen Science: A Driver of Astronomy for Development}
CSR has emerged as a powerful catalyst for astronomy for development, aligning perfectly with the goals of the IAU Office of Astronomy for Development (OAD) in South Africa \footnote{\url{https://astro4dev.org/}}, established to link astronomy with socio-economic development goals \citep{IAU-OAD-Miley-Carignan-Govender2011IAUS}. By involving the public directly in astronomical research, it transforms science from an elite pursuit into a tool for education, empowerment, and social progress \citep{CSR-Bonney2014Science}. CSR initiatives use accessible data and online collaboration to enable students, teachers, and enthusiasts — even in remote or under-resourced regions — to contribute to real discoveries. Projects like RGZ and RAD@home show how open data and mentorship build skills in analysis, reasoning, and digital literacy, strengthening STEM education and capacity building. CSR also promotes gender balance, inclusion, and community engagement, bridging professional research and society while fostering a global culture of curiosity and shared discovery.
By democratizing participation in astronomy, CSR transforms celestial exploration into a socio-economic development tool—inspiring innovation, nurturing talent, and advancing the United Nations’ Sustainable Development Goals (SDGs) through the universal language of the night sky \citep{CSR-SDG-Fritz2019Nature}.

\vspace{-0.2cm}
\section{Citizen Science and Human Rights}
CSR embodies the principles of human rights, particularly the right to participate in and benefit from scientific progress, as stated in Article 27 of the Universal Declaration of Human Rights (1948)\footnote{\url{https://www.un.org/en/about-us/universal-declaration-of-human-rights}}. Article 27 states \textit{``(1) Everyone has the right freely to participate in the cultural life of the community, to enjoy the arts and to share in scientific advancement and its benefits. (2) Everyone has the right to the protection of the moral and material interests resulting from any scientific, literary or artistic production of which he is the author.''} Thus, it declares that access to knowledge and the ability to contribute to discovery should not be limited to professional scientists or wealthy nations, which has been further elaborated by \cite{Shaver2010_RightToScience} \& \cite{Bishop2015_RightToScience}. CSR, as emphasised by UNESCO, promotes open science by enabling people to participate directly in observation, data analysis, and discovery\footnote{\url{https://www.unesco.org/en/open-science}}. This broad access supports equality, inclusion, and freedom of expression, allowing students, teachers, and citizens—regardless of gender, location, or income—to contribute to new knowledge. Such participation strengthens scientific literacy, transparency, and informed decision-making in fields like climate science, public health, and astronomy. It also advances the rights to education and development by building critical thinking and access to global data \citep{VayenaTasioulas2015_HumanRightCitizenScience}. Initiatives like RAD@home show how collaborative research networks can bridge social divides, promote gender equality, and make science widely accessible.

\vspace{-0.2cm}
\section{Considerations and Caveats in Citizen Science Practice}
The success of large online CSR projects, reflected in substantial participation and numerous refereed publications, demonstrates the value of public involvement in astronomical discovery. Nonetheless, this mode of engagement also has recognised limitations that future designs may wish to address. In wide-scale, web-based platforms, volunteer contributions are often structured as rapid classification or identification tasks. This approach is effective for building reliable consensus from many independent assessments, but it can restrict opportunities for deeper analytical involvement or creative reasoning by individuals who may wish to contribute at a more advanced level. Furthermore, because each dataset is examined by many volunteers, individual contributions are necessarily diluted within the wider effort, and personalised recognition—whether through certificates or acknowledgements—does not function as a meaningful academic credential. Co-authorship is possible but typically limited, given the scale and nature of participation. These characteristics are intrinsic to large, distributed platforms, which are optimised for handling enormous data volumes and ensuring statistical robustness, rather than for intensive training or mentorship.

Smaller, community-based CSR initiatives operate differently and therefore address a different set of needs. Programmes such as RAD@home adopt a focused Pro–Am (professional-amateur) collaboratory model in which participants receive structured guidance, interact closely with professional astronomers and engage in more detailed investigation of individual sources. This creates opportunities for university-level students, especially those in regions with limited access to internships or research facilities, to gain meaningful research experience and, in some cases, co-authorship on scientific publications. The depth of engagement achievable in such collaboratories, however, relies on sustained mentorship and is therefore practical only at modest scales. Consequently, these efforts complement rather than replace large web-based platforms: the former prioritise capacity building and participant growth, while the latter excel at harnessing broad public engagement to address large data challenges.

Both models highlight valuable and distinct strengths. Large-scale online platforms provide reach, robustness and efficiency, while collaboratory-style CSR offers depth, training and local impact. Together, they illustrate the diverse ways in which CSR can contribute to radio astronomy and demonstrate how carefully designed frameworks can support both scientific discovery and wider societal benefit in the SKA era.
\section{SKA@home: Proposed CSR with SKA}
CSR already plays a meaningful role in radio astronomy, and the SKA Observatory (SKAO) recognises its value as part of its broader scientific and educational vision. The SKAO outreach portal\footnote{\url{https://www.skao.int/en/resources/outreach-education/citizen-science}} features several active initiatives that engage the public in real scientific work, including projects such as PULSE@Parkes, Radio JOVE, RAD@home, Einstein@Home, RGZ, Gravity Spy, Zooniverse-based programmes, and Bursts from Space using MeerKAT. The SKA construction proposal\footnote{\url{https://skao.canto.global/s/M8159?viewIndex=0}} also highlights existing efforts within member countries, for example, RGZ: LOFAR and RAD@home, which demonstrate the ability of citizen participants to contribute directly to research and discovery. These projects show that well-designed CSR programmes can generate high-quality scientific outcomes, making the SKA era an ideal environment for a dedicated CSR framework.

The Government of India ran a year-long (2019) national outreach programme in four major cities to raise awareness of the SKA, combining exhibitions, public events, and a dedicated SKA week. As part of this effort, RAD@home workshops trained students to interpret GMRT radio images and introduced them to CSR in preparation for future SKA data \citep[]{RAD-Hota2016, RAD-SKAIC-Ramanujam2024}. The programme drew close to six hundred thousand visitors, reflecting strong public interest in large international astronomy projects.

\subsection{SKA@home Discovering “Out-of-the-Box” Objects and “Tip-of-the-Iceberg” Features}
A significant fraction of the most interesting radio sources are neither compact nor textbook double lobes. Interferometric imaging can break extended emission into fragments because of limited sensitivity to large angular scales, while AGNs often undergo multiple duty cycles that leave behind faint, irregular relic lobes. In many cases, only a small part of the true structure is obvious at first glance; the rest sits just above the noise, or is spread across multiple images, resolutions, and surveys. Recognising that several apparently unrelated patches of emission belong to one physical system requires a global view of the field, an ability to ignore confusing foreground and background sources, and a willingness to test unconventional interpretations. This kind of holistic pattern recognition is still very hard to encode in automated pipelines.

Episodic or DDRGs are a clear example. Their outer/relic lobes are often diffuse, asymmetric, or partially blended, so that the full episodic structure becomes evident only when deep low \& high frequency maps are examined together \citep[e.g.,][]{DDRG-review-Saikia2009BASI, Speca-Hota2011, Speca-likeSethi2025, giant-DDRG-LoTSS-Dabhade2025}. For example, extended relic emission in systems such as Speca is visible in low-resolution NVSS radio surveys but disappears in higher-resolution FIRST (Fig.\,\ref{fig:example_figure}, middle panel). Its true nature can only be understood by combining information from multiple telescopes for structure and spectral index \citep{Speca-Hota2011}. Similar challenges appear in galaxy groups and clusters, where old radio bubbles and relic plasma can extend over hundreds of kiloparsecs with very low surface brightness. The complex system of multiple distorted bubbles reported by \citet{Brienza-LOFAR-relic-bubbles-2021Nature} illustrates how difficult it can be to trace a sequence of activity episodes across such a wide range of spatial scales. In many of these cases, the most informative emission is close to the detection limit or embedded in other structures, so that only careful visual exploration reveals the underlying story. Multi-telescope multi-frequency follow-up is needed to determine the synchrotron spectral age of the relic \citep[e.g.,][]{DDRG-aging-Konar2006} and similarly UV-optical-IR follow-up and spectroscopy are needed to determine the star formation history. A sample of such old relic radio emission, found around post-starburst or post-merger galaxies, which could be as old as the time-scale of quenching of star formation, is still missing. If found by SKA, it would be "smoking gun" evidence of AGN feedback models \citep[e.g.,][]{NGC3801-Hota2012, RAD-Hota2016}. This is in line with one of the eight science goals of SKA, understanding cosmic evolution of star formation and AGN activity. For a review on feeding and feedback on galactic scale, an article by \cite{Maccagni01.2026.SKA} can be found in this volume.

The challenge becomes even greater for genuinely new classes of objects that have no clear precedent in existing catalogues (e.g. RAD-ORCs and RAD-BAARG). Their enormous, diffuse ring-shaped emission was not predicted by standard models of radio galaxy morphology, and the features did not stand out immediately in individual surveys. It was only by examining their structure across multiple frequencies and by considering their environments that their unusual nature became evident \citep{RAD-ORC-Hota2025MNRAS, RAD-BAARG-Hota2026MNRASL}. ML methods perform well when trained on known morphologies, but they are intrinsically less reliable for rare sources that lie outside the distribution of their training sets. In contrast, a trained human eye can ask whether something looks out of place, incomplete, or only partly revealed, and can make use of heterogeneous archival data to test that impression. Likely, when the narrow twin ring structure of the ORC is blended with larger-scale diffuse radio emission, AI/ML may confuse, but a trained human eye spots it easily. Similarly, narrow structures embedded inside large-scale diffuse emission also require careful visual multi-telescope analysis to interpret the evolution of the AGN activity and merger history of the host galaxy (e.g., horse-shoe shape by \cite{Horse-show-Kumari2024}; S-shape by \cite{S-shaped-Misra2025}, mini-double by \cite{halo-DDRG-Jamrozy2007}.

Automated anomaly detection will play an important role in managing the vast data volumes expected from the SKA, and several techniques are already being developed for this purpose \citep{Astronomaly-Lochner2021}. However, the most interesting candidates produced by these algorithms still require human inspection, especially when a multi-wavelength context is essential. For example, the peculiarity of a one-sided radio jet in RAD12 was not so intriguing until the photometric redshift of the neighbouring galaxy, where the extended radio emission abruptly stops (Fig.~\ref{fig:example_figure}, right panel), was found to be similar to the host of the radio jet \citep{RAD12-Hota2022}. Inspired by that, RAD@home citizen scientists have found several new cases of jet-galaxy interactions from radio-optical structural comparisons incorporating redshift information in optical databases.     

These considerations argue strongly for keeping human inspection in the SKA discovery loop. Anomaly detection algorithms can narrow down the search space by flagging unusual or low-confidence cases, but the final step of deciding whether a fragmented, low-signal feature is in fact the visible tip of a much larger and more complex system will often require human judgment. CSR programmes such as SKA@home can possibly provide a natural way to mobilise this capacity at scale. With targeted training in radio morphology and access to multi-survey visualisation tools, volunteers can help to connect apparently disjoint components, identify out-of-the-box sources, and reveal the true iceberg beneath the visible tip. In doing so, they will not only complement automated techniques but also open a path to discoveries that current algorithms are not yet designed to find.
  
\subsection{A Practical Framework for SKA@home}
We envisage SKA@home as a flexible, globally accessible CSR programme that complements SKA surveys while supporting education and capacity building. 

\begin{enumerate}
    \item \textbf{SKA@home web-based:} A small, carefully curated fraction of SKA continuum survey data, the ``Discovery Stream'', would be released through an online platform modelled on successful CSR projects such as RGZ. These data would be pre-processed and optimised for public classification, enabling volunteers to identify radio morphologies, flag anomalies, and highlight unusual or rare sources.

\begin{itemize}
    \item Participation would be open worldwide through a unified certification portal. Short, structured online tutorials would guide volunteers from basic to advanced levels, ensuring consistent data quality without creating barriers to entry. This global approach avoids national gatekeeping and allows individuals from all regions, including those with limited access to astronomical infrastructure, to contribute meaningfully.

    \item The analysis pipeline would combine CSR with ML in a hybrid workflow. Automated classifiers would handle the majority of routine detections, while volunteers focus on cases where algorithms disagree, where uncertainties are high, or where unexpected structures appear. This maximises scientific return while ensuring that human insight remains central to identifying complex or rare phenomena.
\end{itemize}

\item \textbf{SKA@home Collaboratory:} A complementary approach, inspired by existing community-driven programmes (e.g., RAD@home), is to provide a very small and well-curated subset of SKA continuum survey data for use in a structured CSR framework. Registered CSR groups, recognised by SKAO on the basis of demonstrated scientific engagement and community benefit, would receive access to these data together with training resources and workshop material delivered through national EPO activities. 

\begin{itemize}
    \item Trained participants could then examine selected images under the guidance of local scientists and submit First Investigation Reports (FIRs) to a centralised SKA@home database. Local institutions from all participating nations may independently organise CSR workshops to strengthen engagement, but all participants interact with the same central FIR submission web-portal at SKAO. Thus, the database would automatically prevent duplication based on source position and extent, while maintaining transparent provenance for all submissions. After a defined proprietary period, FIRs would become visible to scientists across all participating countries for further study and potential follow-up, ensuring both credit and scientific continuity.

    \item This Collaboratory model allows citizen scientists to contribute structured early-stage analysis while enabling professional researchers to evaluate and develop promising FIRs into full scientific results. By combining training, mentorship, and a clear pathway from public participation to scientific output, such a scheme can support both research and wider societal engagement in the SKA era.
\end{itemize}

\end{enumerate}

Contributions from volunteers would be acknowledged in line with standard practices used across large CSR initiatives. Publications based on SKA@home classifications or discoveries would credit the SKA@home programme and recognise volunteers who made substantial contributions, while the scientific analysis, interpretation, and authorship remain the responsibility of the professional research teams. This model offers a transparent and equitable pathway for public participation to feed into peer-reviewed science, without creating ambiguity around scientific responsibility or authorship.

\textbf{Dedicated follow-up of SKA@home discoveries}: 
To maximise SKA@home’s scientific value, SKAO could establish a streamlined mechanism for limited, rapid follow-up of high-priority objects flagged by volunteers. Similar precedents exist, such as GMRT follow-up of RAD@home discoveries through GOOD-RAC \citep{RAD12-Hota2022} and targeted HST observations of Zooniverse systems in the GEMS of the Galaxy Zoos \citep{ZooGEMS-Keel2022}. Any SKAO process would align with existing time-allocation rules but include provisions for modest, timely observations where CSR classifications show strong potential. Such a pathway would boost the impact of CSR-driven discoveries and inspire future astronomers, forming part of a scalable, inclusive SKA@home framework supported by training, AI, and global community participation.
               
\subsection{Prospects for a distributed network of SKA-inspired low-frequency stations}
In the long term, once SKAO operations are established, a global network of SKA-inspired low-frequency stations could be developed by partner institutions, especially in remote or underdeveloped regions. These would be simplified arrays meeting basic standards for timing and calibration. Their primary role would be education and capacity building, enabling local studies of the ionosphere, space weather, meteor trails, and lightning-related radio emission. With defined protocols, selected data could also support SKA campaigns in heliospheric science, transients, and VLBI, extending coverage beyond Australia and Africa. A pilot project to assess feasibility would be required before its realisation. 

 \vspace{-0.2cm}
\section{Conclusions} 
Over the past two decades, astronomical publications using CSR have grown steadily, now exceeding 100 per year in NASA ADS. Millions of participants worldwide have contributed to such initiatives, which not only generate high-quality scientific results but also deliver educational and societal benefits. In this paper, we reviewed major CSR platforms in astronomy, with emphasis on RGZ, RGZ–LOFAR, and the RAD@home Astronomy Collaboratory. While web-based projects like RGZ excel in large-scale catalogue generation, RAD@home advances the CSR model by providing direct, structured training that enables participants to develop into semi-professional e-astronomers capable of discovering rare and complex radio sources.

Aligned with the goals of the IAU’s Office of Astronomy for Development and the United Nations’ recognition of the right to participate in science, we strongly advocate the incorporation of CSR within the SKAO. We propose two complementary operational modes: SKA@home-web-based, a fully online system for large-volume classification, and SKA@home-Collaboratory, a training-driven model that builds student communities capable of submitting preliminary discovery reports to SKAO. Each participating nation would oversee the training pathway and publication process. Through these initiatives, the SKA could connect meaningfully with university students and astronomy enthusiasts across the globe, fostering widespread scientific engagement and discovery.

\vspace{-0.5cm}
\section{Acknowledgement}
\vspace{-0.25cm}
AH acknowledges the University Grants Commission (UGC, Ministry of Education, Govt. of India) for his monthly salary grants since June 2014. We sincerely thank Late Prof. Govind Swarup, FRS, for his encouragement during the formative years of the RAD@home Collaboratory. We also gratefully acknowledge the significant contributions of Late Dr. Nandivada Rathnasree, Ms. Megha Rajoria, Mr. Avinash Kumar, and Mr. Ck. Avinash, whose efforts through various camps and workshops greatly contributed to the growth of RAD@home. 

\vspace{-0.5cm}
\bibliographystyle{abbrvnat-maxbibnames4}

\bibliography{chapter} 

\end{document}